\documentclass[12pt]{article}
\usepackage{amssymb,amsmath}
\usepackage{euscript}
\def\lie{{\EuScript L}_f}
\def\mystar{\,{}^{\overline *}}
\def\hodge{\,{}^*}

\begin{document}
\date{}

\title{Covariant formulation
of the Liouville \\ and the Vlasov Equations}

\author{{O.I. Drivotin}\\
{\small St.~Petersburg State University, St.~Petersburg,  Russia}\\
{\it\small  e-mail: drivotin@ya.ru}
}

\maketitle
\begin{abstract}
  
The fundamental concept of phase space
for particles moving in the four-dimensional spacetime
is analyzed.
Particle distribution density
is defined as differential form,
which degree may be different in various cases
It should be emphasized that this approach
does not include the concepts of phase volume and distribution
function.
The Liouville and the Vlasov equations are written
in tensor form.
The presented approach is valid in both
non-relativistic and  relativistic cases.
It allows using arbitrary systems of coordinates
for description of the particle distribution.
In some cases, making use of special coordinates gives possibility
to construct analytical solutions.
Besides, such approach is convenient
for description of degenerate distributions,
for example, of the Kapchinsky-Vladimirsky distribution, 
which is well-known in the theory of charged particle beams.
It can be also applied for description of mass distributions
in curved spacetime.

\end{abstract}

 

\section{ Introduction }

Covariant approach is commonly in use in general relativity.
But it can be also applied in all physical theories
where processes in space and time are considered.
Covariance means 
using of  objects that can be defined without introducing coordinates,
for example, tensors.
Covariant tensor equality
is not associated with any coordinate chart.
Nevertheless,
if coordinates are specified,
a tensor equality can be regarded as
a set of equalities of corresponding tensor components.

 Customary forms of the Liouville and the Vlasov equations
 contain  partial derivatives 
 with respect to space coordinates and components of the momentum vector
  \cite{lib,groot}. 
Therefore, they cannot be regarded as covariant,
because differentiating
with respect to vector components is not covariant operation.
Covariant approach allows 
making use such operation as gradient of function,
which components are derivatives 
with respect to coordinates in some space.
But in general case,
components of the momentum vector cannot be taken
as global coordinates in a phase space,
as it is explained in Sec.2.

In all of these works,
the concept of distribution function was employed.
The definition of a distribution function
requires making use the concept of phase volume.
As distinct from the works cited above,
presented here approach does not include 
the concepts of distribution function and phase volume.
The particle distribution density
is introduced by a natural way
as a differential form of some degree.
The main ideas of the present  work 
were previously formulated in Refs.
\cite{drivpac2011,drivrupac2012}.

Kinetic equations in covariant form
were also  written down in some  works
(see, for example, Refs. \cite{holm,holm2008,plasmas2013}).
But  forms of  top degree only \cite{holm,plasmas2013})
and particle distribution function \cite{holm2008,plasmas2013}
were employed.

The presented here approach is valid
both for nonrelativistic and for relativistic cases equally.
The difference between these cases
relates only to the form of dynamics equations.
In classical problems covariant approach
has also great importance.
It allows making use of various coordinates in the phase space.
It is required, for example,
in problems concerned with
self-consistent particle distributions
for a charged particle beam.
Besides, this approach allows to consider degenerate distributions
that are described by forms of lower degree. 

In fact, such approach was used in numerous
works devoted to self-consistent distributions
for a cylindrical beam and for a beam with varying cross-section radius
("breathing" beam) 
[8-13]
.
In these works, the space of  motion integrals was introduced,
and particle density was defined in this space.
Under some conditions,
the phase density can be expressed through
the introduced density.
It was convenient to consider the phase density
as a differential form, and the motion integrals  as phase coordinates.
It gave possibility
to construct new solutions of the Vlasov equation.
For example,  they can be obtained as  a linear combinations
of degenerate distributions, or  as  solutions of some integral equation
[9-13].

To verify that these distributions
are solutions of the Vlasov equation,
we can take into account that the phase density conserves 
along the characteristic lines of this equation.
To provide rigorous proof,
we need an equation where one can substitute these distribution.
One of the purposes of this article is to write down such equation.
Therefore,
formalization of the approach
applied in works
 [8-13]
is offered in this article.

The article is organized as follows.
In the second section the  concept
of phase space is introduced.
The third section is devoted
to a definition of phase density.
The fourth section deals with the Liouville and the Vlasov equations,
which are written in covariant form.  
The difference between these two equations
consists only in  a way of computation of the force
acting on the particles.
In the Liouville equation the force is external one,
while in the Vlasov equation it should be computed
using a particle distribution the equation is written for.
In the fifth section
the case of electric charged matter is considered.
Relation between the phase density
and the four-dimensional current density 
entering the Maxwell equations is obtained.
As the conception of the 
four-dimensional current density
is applicable both in relativistic as in nonrelativistic cases,
this relation is valid in both cases.
In the sixth section two known
examples are considered to show how new approach works.
They are the Brillouin flow \cite{brill}
and the Kapchinsky-Vladimirsky distribution \cite{kapmono}.
The seventh section contains short discussion of the results.


\section{Phase Space}

Physical processes occur in the four-dimensional spacetime.
Description of motion of a particle
requires specifying not only its position in spacetime,
but also its velocity.
To identify state of a particle with account of its velocity  
we should define a phase space.
When a phase space is defined,
state of a particle can be regarded as 
a point in it.

Consider an open domain $D$ in the spacetime.
Assume that there exists 
some foliation formed by disjoint spacelike surfaces filling $D.$
Let's parameterize these surfaces
with a continuous parameter $t.$

Assume also that there exists
a system of diffeomorphic mappings
of these surfaces to some selected surface.
Call this selected surface
the configuration space and denote it by $C.$
The assumption introduced above means 
that the domain $D$ can be represented 
as $T\times C,$
where $T$ denotes an interval where the parameter $t$ varies, $t\in T.$ 
Such approach is widely used in general 
relativity, and it is called $3+1$ splitting of the spacetime
\cite{adm,dirac,gour}.

When time passes,
particles move
from one surface to another.
One can describe dynamics of particles
as  dynamics of their images in the configuration space.
Call the tangent bundle \cite{god}
of the configuration space the phase space,
and denote it by $M,$  ${\rm dim}M=6.$
Points in $M$ will be denoted by $q,$ $q\in M.$
 
Such phase space  is senseless from the physical point of view, but it can be used.
In order to construct a meaningful phase space,
let's apply the fundamental physical concept of reference frame.
Following to  Ref. \cite{refframes,refframes2},
let us introduce a congruence of observers in $D.$
Call one of them  the main observer and call the others the local observers.
Assume that the main observer can measure the time in $D.$
Assume that at any instant of time measured by the main observer
one can find a surface such that all events of it are simultaneous
from the point of view of all observers of the congruence.
Call such surface the layer of simultaneous events.
It is easy to see that the layers of simultaneous events
form a foliation in $D.$

Assume that there exists
a system of diffeomorphic mappings
of these layers to one selected layer
such that for any event  $P$ its image is
the intersection of the selected layer
with worldline of an observer from the congruence
that passes through $P.$
The selected layer will be also called the configuration space, as previously. 

The congruence of the observers and the system of the introduced mapping
form a reference frame.
It can be shown that  coordinates of image of an event
in the configuration space can be regarded as spatial coordinates of the event
and time measured by the main observer is  temporal coordinate of the event
\cite{refframes,refframes2}.
As previously, call the tangent bundle of the configuration space 

the phase space.
Note that this definition is valid
both in classical and relativistic cases.

Phase space associated with a reference frame is preferable,
because all values entering equations describing physical processes
have physical sense.
For example,
in temporal-spatial coordinates
associated with a reference frame
components of the electromagnetic field tensor 
are components of the electric field and components of the magnetic field.
Further we shall consider only such phase space.

According to definition of the tangent bundle \cite{god},
for every $x\in C$
there exists a neighborhood $U$
and a homeomorphic mapping $\Phi:$ $p^{-1}(U)\mapsto U\times T_xU$
such that $p_1\Phi=p.$
Here $T_xU$ denotes the tangent space at the point $x,$
which is fiber of the bundle,
$p$ and $p_1$ denote projections of $M$ and $U\times T_xU$ on $U$ 
correspondingly.
It is clear that if one take three spatial coordinates $x^1,$ $x^2,$ $x^3,$
then corresponding components of velocities can be taken
as coordinates in $T_xU,$
and these 6 numbers can be taken as coordinates locally
 in $p^{-1}(U)$.

Global coordinates in the phase space can be introduced as follows.
Consider a subregion in the phase space 
corresponding to some subregion in the configuration space,
which will be denoted also by $C,$
instead of the whole phase space.
It allows  taking into account cases
when spatial coordinates of some coordinate chart
are defined not in all configuration space.

Assume that $C$ is simply connected.
Specify coordinates in $C,$
and assume that components of the metric tensor
in these coordinates are continuously differentiable in $C.$

Take some point $x_0\in C.$
Consider the boundary problem
\equation
\frac{\partial v^i}{\partial x^j}+\Gamma^i_{jk}v^k=0,\quad v(x_0)=v_0
\label{eq:par}
\endequation
where $v_0$ is some vector, $v_0\in T_{x_0}C,$
and $\Gamma^i_{jk}$ denotes the Christoffel symbol of the second kind.
Here and further we use the Einstein summation convention
according to which summation is meant
over all allowed values
of repeated upper and lower indices.

The equation of parallel transport of a vector $v$
along a line $x=x(\lambda)$
can be written in the form 
$$
\frac{d v^i}{d\lambda}+\Gamma^i_{jk}v^k\frac{dx^j}{d\lambda}=
(\frac{\partial v^i}{\partial x^j}+\Gamma^i_{jk}v^k)\frac{dx^j}{d\lambda}=0.
$$
Therefore, the problem (\ref{eq:par})
describes a vector field
such that any vector of this field at a point $x$
can be regarded as a result
of parallel transport of the vector $v_0$
along any line connecting points $x$ and $x_0.$
 
This boundary problem
for the system of linear differential equations has unique solution
for simply connected region.
Indeed, solution of the problem (\ref{eq:par})
consists in finding of 3-dimensional surfaces in the phase space
which tangent vectors are annihilated by forms
$\omega_{(i)}=dv^i+\Gamma^i_{jk}v^k\,dx^j,$
$i=\overline{1,n}.$
It is sufficient for solvability that all $d\omega_{(i)}=0.$
Differentiating $\omega_{(i)},$
we get 
$d\omega_{(i)}=\Gamma^i_{jk}dv^k\wedge dx^j=0,$
as Christoffel symbols are permutation symmetric.

Therefore,
for each vector $v\in T_xC$
we can find parallel vector $P_{x_0}v\in T_{x_0}C$
at the point $x_0.$
Mapping $v\mapsto P_{x_0}v$
is continuously differentiable,
as it is set by the Green functions of the problem (\ref{eq:par}).
Then three coordinates in the configuration space
and three components of $P_{x_0}v$
in these coordinates
can be taken as coordinates 
of the point of the phase space
specified by $x$ and $v.$
Besides, continuous differentiability of the mapping ensures
that transition functions between various coordinates are differentiable.

It means that phase space $M$ is differentiable manifold.
 

If particles always lie on the same surface in the phase space,
or distribution density does not depend on some coordinate,
then the phase space $M$ can be taken as corresponding subspace
of the initial phase space. 
In this case dimension of the phase space under consideration
is less then dimension of the initial phase space,
and coordinates are not necessarily 
spatial coordinates and velocity coordinates
and can represent a mixture of both kinds.

\section{Phase Density}
In this section we shall concern various types of distributions,
all of which can be described on the base of a common approach.

As a simplest case, consider continuous media
that occupies an open set in the phase space.
Within the framework of this model,
particle number in an open subregion $G,$
$G\subset M,$
is not necessarily integer. 
Call the differential form 
 $n(t,q)$   of degree $m=\dim M$
such that integration of the form over 
each  open set $G$  
gives  particle number in $G$
the particle distribution density in the phase space, or the phase density:
\equation
\int\limits_Gn=N_G.
\label{eq:volumepart}
\endequation
The boundaries  of $G$ and the form $n$ are assumed sufficiently smooth
for integration being possible.

Consider also another case
when particle are distributed on an oriented surface $S$
in the phase space
that can move, $\dim S=p,$ $0<p<m.$
Call the differential form $n(t,q)$ of degree $p$ 
defined on the surface $S$
such that for any open set $G,$ $G\subset M,$
\equation
\int\limits_{G\cap S}n=N_G
\label{eq:surfacepart}
\endequation
the particle distribution density for this case.
This form depends on orientation of the surface,
which is defined by an ordered set of $m-p$ vectors.
A change of the orientation can result in change of sign of 
the form components \cite{drivmono}.
Assume that form $n$ and the surface $S$ are also sufficiently smooth
for integration being possible.

At last, consider the case of ensemble of pointlike particles.
Define the scalar function
\equation
\delta_{q'}(q)=
\begin{cases}
1,\qquad q=q',\cr
0, \qquad q\neq q'.
\end{cases}
\label{eq:delta}
\endequation
If  $q'$ depends on $t,$
then this function is also function of $t.$
All functions 
which values are nonzero  
only in finite set of points 
can be represented as linear combination of the functions of form (\ref{eq:delta}).
Restrict ourselves only to combinations with all coefficients equal to $1:$  
\equation
n(t,q)=\sum\limits_{i=1}^N\delta_{q_{(i)}}(q),\qquad q_{(i)}\neq q_{(j)},\quad {\rm if}\quad i\neq j.
\label{eq:sumdelta}
\endequation
In this class of functions,
define an operation of taking sum of function values
in all points $q_{(i)}$ where the function value is nonzero:
\equation
\sum\limits_{q\in G} n(t,q)\equiv\sum\limits_{i:\, q_{(i)}\in G} n(t,q_{(i)}).
\label{eq:intdelta}
\endequation
Operation defined by equation (\ref{eq:intdelta})
is analogous to integration of the form of higher degree over $G.$
A scalar function can be regarded as the differential form
of degree $0.$
Therefore, equation (\ref{eq:intdelta}) 
set a rule of integration of a form of degree $0$ over open set $G.$

As previously,
call  function of form (\ref{eq:sumdelta})  
the phase density for system of pointlike particles
if
\equation
\sum\limits_{q\in G} n(t,q)=N_G.
\label{eq:pointpart}
\endequation
It is easy to understand that
the phase density is given by equality (\ref{eq:sumdelta}),
where  $q_{(i)}$ are positions of the particles in the phase space,
$i=\overline{1,N},$
$N$ is the total number of particles in the ensemble.

These three cases  can be combined as follows.
Denote by $G_0$ a region in the phase space 
enclosing all open sets $G$
which can be considered.
Consider a linear space $\cal F$
of some integrable test functions defined on the set $G_0.$  
Define functional $<n,f>$ as 
$$
\int\limits_{G_0}n(t,q)f(q),\quad \int\limits_{G_0\cap S}n(t,q)f(q),
\quad \sum\limits_{q\in G_0}n(t,q)f(q)
$$
in the first, the second, and the third cases 
correspondingly, $f\in{\cal F}.$
Then definition of the phase density $n$
can be written in the form
\equation
<n,\chi_G>=N_G,
\label{eq:defphaseden}
\endequation
where $\chi_G$ is the characteristic function
of the set $G,$ $G\subset G_0:$
$$
\chi_G(q)=
\begin{cases}
1,\quad q\in G, \cr
0,\quad q\notin G.
\end{cases}
$$

\section{The Liouville and the Vlasov Equations}

At each instant of time $t\in T$,
particle dynamics equations define a vector field $f(t,q)$
in the phase space.
Assume that for each $t\in T,$ $q\in M$
there exists a unique integral line passing through point $q.$
For example, if components of $f(t,q)$ in Cartesian coordinates 
are continuously differentiable with respect to coordinates and the time,
it will be so. 
The time can be taken as a parameter for the integral lines.

We shall use operation of the Lie dragging \cite{drivmono}
to describe how the phase density changes when the time passes.
Lie dragging  of a point $q$ 
is the point gotten by displacement of  $q$
along the integral curve passing through $q$
by the parameter increment $\delta t.$
Denote it by $F_{f,\delta t}q.$
By virtue of uniqueness of the integral
curve passing through a point,
this mapping is reversible.
 
The mapping  $F_{f,\delta t}q$
induces the following coordinate transformation.
For each point $q,$
let's take coordinates of its preimage.
Such transformation can be regarded as shift of the system of coordinates.
Denote these coordinates as $q^1_{(f,\delta t)},$\ldots, $q^6_{(f,\delta t)}.$

Let  some tensor field $T$ be defined in the phase space.
Define tensor at the point $F_{f,\delta t}q$ as follows:
its components in coordinates $q^i_{(f,\delta t)},$
are equal to components of $T$ at the point $q$ in the initial coordinates.
Such tensor is called the Lie dragging of the tensor $T$
along vector field $f$ by the parameter increment $\delta t.$
Denote this tensor by $F_{f,\delta t}q.$

How do phase density changes when particles move?
It is easy to understand that degree
of the differential form describing particle distribution does not change,
because we can take
dragged  vectors of basis at the initial point $q$
as the  basis vectors at the dragged point $F_{f,\delta t}q.$
Assume also that the particles do not appear and disappear.
Therefore,
integrals of the phase density 
over any set should be the same as integral over dragged set.
For example, for a form $n$ of top degree $m$
$$
\int\limits_{F_{f,\delta t}G}n(t+\delta t,q)=\int\limits_G n(t).
$$
Here $F_{f,\delta t}G$ denotes an image of the set $G,$
that is set all points of which are images of points of $G.$
Introducing coordinates $q^i_{(f,\delta t)}$
we see that $F_{f,\delta t}G$ looks in these coordinates as $D$
in coordinates $q^i.$
Therefore, to ensure equality of the integrals
for any region it is necessary and sufficient that
the following equation will be satisfied
\equation
n(t+\delta t,F_{f\,\delta t}q)=F_{f,\,\delta t}n(t,q).
\label{eq:master}
\endequation

It is easy to see that
analogous reasons take place
when degree of form $n$ is less than $m.$
Therefore, equation (\ref{eq:master})
is valid in these cases also.

Vector field $f$ depends on the force acting on a particle.
Let us call equation (\ref{eq:master}) the covariant form of the Liouville equation
if the force is purely external.
Let us call equation (\ref{eq:master}) the covariant form of the Vlasov equation
if the force is determined  also by a self field,
by which we mean field produced by the particle ensemble.
In the first case, the equation (\ref{eq:master}) 
can be regarded as a differential equation.
Though it does not contain derivatives,
it describes such object as differential form.
In the second case, the equation (\ref{eq:master}) 
can be regarded as an integro-differential equation,
because usually it is possible
to represent the self field 
in an integral form with use
of the Green functions.  
We assume here that
structure of separate particles can be ignored,
and one should take into account only phase density
while computing the force acting from the self field.
Account of a particle structure 
gives an additional term in the Vlasov equation
which is called the collision integral.
This term is sufficient only at high densities,
and it will be omitted from further consideration here.

Consider the case when the particle distribution is described by
a top degree form.
How does this form change at some point $q$
of the phase space depending on the time?
Assume that its only component $\tilde n$ 
is continuously differentiable with respect to phase coordinates and time.

Let at some instance of time $t$
the phase density at a point $q$
is equal to $n(t,q).$
At the instance $t+\delta t$
it will be equal to
$n(t+\delta t,q)=F_{f\,\delta t}n(t,F_{f\,-\delta t}q),$
as the phase density changes according to equation  (\ref{eq:master}).
Introduce the derivative of a differential form with respect to the parameter $t$
as a form which components are derivatives of corresponding components
with respect to $t.$
Then we obtain the Liouville and the Vlasov equation in the form 
\equation 
\frac{\partial n}{\partial t}=
\lim\limits_{\delta t\rightarrow 0}
\frac{n(t+\delta t,q)-n(t,q)}{\delta t}=
-\lie n(t,q).
\label{eq:master6}
\endequation
Here $\lie n(t,q)$ denotes the Lie derivative of the phase density
along the vector field $f,$
which can be defined  as follows.
The Lie derivative of a tensor field $T$ along a vector field $f$  is
\equation
\lie T=\lim\limits_{\delta t\rightarrow 0}
\frac{T-F_{f,\,\delta t}T}{\delta t}.
\label{eq:lee}
\endequation

The equation (\ref{eq:master6})
for density form
of top degree was considered 
also in   works   \cite{holm,plasmas2013}.
For this case, it is easy to understand 
that the solution of (\ref{eq:master6})
  is also satisfies to the equation (\ref{eq:master}) \cite{holm}.

Components of the Lie derivative of a differential form $T$
of   degree $p$   can be determined from the equalities
\equation
(\lie T)_{i_1\ldots i_p}=
\frac{\partial T_{i_1\ldots i_p}}{\partial q^k}
f^k
+\frac{\partial f^j}{\partial q^{i_1}}
\cdot
T_{j\,i_2\ldots i_p} +
\ldots
\frac{\partial   f^{j}}{\partial q^{i_{p}}}\cdot
T_{i_1\ldots i_{p-1}\,j}.
\label{eq:leeform}
\endequation
Summation is meant
over possible values of repeated  indices.

When the phase space is associated with a reference frame,
the dynamics equations can be written in the form \cite{drivmono}
\equation
\frac{dx^i}{dt}=v^i,\quad m \sum_{j=1}^3g_{ij}(\frac{d}{dt}(\gamma v))^j =Q_i,\, i=1,2,3.
\label{eq:partdynamics2}
\endequation
Here $g_{ij}$ and $(d/dt(\gamma v))^j$ 
are spatial components of the metric tensor
and of covariant derivative
of vector $\gamma v$ correspondingly,
$\gamma$ is reduce energy of a particle,
$Q$ denotes three-dimensional   force vector 
acting on a particle.

For example, if particles move in the electromagnetic field,
then 
\equation
Q_i=  e(E_i+\sum_{j=1}^3v^jB_{ij}).
\label{eq:emforce}
\endequation
Here $e$ and $m$ are electric charge and mass of a particle,
$E$ is the electric field intensity,
and $B$ is the magnetic   flux density,
$B_{ij}= \mu_0g^{-1/2}g_{il}g_{jm}\varepsilon^{klm}H_k$
\cite{drivmono},
$g$ is the determinant of the spatial part of the metric tensor,
$\varepsilon^{klm}$ are the three-dimensional Levi-Civita symbols,
$H$ is the magnetic field intensity,
$\mu_0$ is the magnetic constant.

As an example, consider ensemble of nonrelativistic particles
in the flat spacetime
which dynamics is described by equation
(\ref{eq:partdynamics2}) with force term 
(\ref{eq:emforce}).
Take Cartesian coordinates
in the configuration space and Cartesian components of velocities
as coordinates in the phase space.
According to (\ref{eq:emforce})
components of the force acting on a particle does not depend
on corresponding components of velocities.
Then in the right hand side of equality (\ref{eq:leeform})
one should take into account only the first term,
and equation (\ref{eq:master6}) takes the form
\equation
\frac{\partial\tilde n}{\partial t}+\sum\limits_{i=1}^3 v^i\frac{\partial \tilde n}{\partial x^i}+
\sum\limits_{i=1}^3 \frac em 
(E_i+\sum\limits_{i=1}^3B_{ij}v^j)\frac{\partial \tilde n}{\partial v^i}=0.
\label{eq:tradvlas}
\endequation
This form of the Vlasov equation is widely used 
in applied problems, for example,
in the theory of   charged particle beams.

The simplest particular case
is an ensemble consisting of one particle.
In this case,
the particle density
is $\delta_{q_{(1)}(t)}(q).$
This scalar function is equal to $1$
only at the point where the particle is located at the instant of time $t,$
and is equal to $0$ at other points.
It is easy to understand that this density satisfies to the equation (\ref{eq:master}).

Density for a  system of $N$ particles
taken in the form  (\ref{eq:sumdelta})
also satisfies to the equation (\ref{eq:master}),
which can be regarded as the Liouville equation
if particle interaction is neglected and
the Vlasov equation otherwise. 

\section{Computation of Self Electromagnetic Field}

The electromagnetic field can be described by
the tensor of electromagnetic field $F.$
When the first coordinate is temporal,
for example, $x^0=ct,$
and the other three are spatial,
components of the tensor are
$$
\|F_{ik}\|=
\begin{pmatrix}
0&E_1/c&E_2/c&E_3/c\cr
-E_1/c&0&-B_{12}&-B_{13}\cr
-E_2/c&B_{12}&0&-B_{23}\cr
-E_3/c&B_{13}&B_{23}&0
\end{pmatrix}
.
$$
 
For simplicity,
consider the case
when the source of
the electromagnetic field 
is electrically charged matter 
describing by
the differential form $J$ of the third degree
in the four-dimensional space-time,
the integral of which over smooth
oriented three-dimensional surface $S$  
gives the quantity of charge passing through $S$ in the direction of its orientation:  
$$
\int_SJ=Q_S.
$$
This form is known as the current density form \cite{drivmono}.
In the coordinates associated with a reference frame,
spatial component $J_{123}$ is the charge density,
and the other components are equal
to components  
of the current density form of the second degree
in the three-dimensional configuration space
with an accuracy up to the sign.

Then the Maxwell equations can be written
in the form \cite{drivmono}
\equation
\begin{cases}
dF=0,\cr
d\mystar F=J/(\varepsilon_0 c)
\end{cases}
\label{eq:farad}
\endequation
where $\mystar$ denotes 
sequential application of
the Hodge operator denoted as ${}^*$ and lowering  of indices
with the four-dimensional metric tensor: 
$$
(\mystar F)_{ik}=g_{il}g_{km}(\hodge F)^{lm},\qquad
(\hodge F)^{lm}=|g|^{-1/2}\varepsilon^{ijlm}F_{ij}, 
$$
$g$ is determinant of the metric tensor,
$\varepsilon^{ijlm}$ is the four-dimensional  Levi-Civita symbol,
$\varepsilon_0$ is electric constant.

In what follows,
we shall concern ourselves
how to express the current density $J$ through the phase density $n.$
For simplicity, 
assume that the phase density is also described by the form of
top degree, $6.$

Consider a simple connected subregion $C$ 
of the configuration space.
According to the previous,
in the phase space, which is tangent bundle of $C,$
one can introduce six coordinates,
three of them being coordinates in $C.$
It means that they are spatial coordinates.
Another three coordinates can be 
expressed through these spatial coordinates and components 
of velocity at this point,
and, conversely,
velocity at some point can be expressed
through these six coordinates.

Denote the spatial coordinates by $x^i,$
$x^i=q^i,$ $i=1,2,3.$
Assume that the particles occupy
some region in the phase space.
Denote the section of this region
at the point $\{x^1,x^2,x^3\}$
by $\Omega(x^1,x^2,x^3).$
Let us call $\Omega(x^1,x^2,x^3)$
the set of admissible values
of the phase coordinates $q^4,q^5,q^6$
at the point of $C$ with coordinates $x^1,x^2,x^3$
A value of particle velocity
is admissible only
if it corresponds to some  point from $\Omega(x^1,x^2,x^3)$.
  
Integrating the current density form
over $\Omega(x^1,x^2,x^3)$
  we obtain the density in the configuration space
\equation
J_{123}(x)
=\int\limits_{\Omega(x^1,x^2,x^3)}
n_{123456}(t,q)\,dq^4\wedge dq^5\wedge dq^6.
\label{eq:J123}
\endequation

In order to find another components of the current density,
for example, $J_{012},$
we take a point $x$ in the spacetime,
and consider a small 3-dimensional cell
spanned by the edge vectors 
$\delta x^0e_{(0)},$ $\delta x^1e_{(1)},$ $\delta x^2e_{(2)}.$
Here $e_{(i)}$ 
denotes basic vector of the coordinate basis.
The cell contains points with coordinates
$ x^0+\alpha \delta x^0,$ 
$ x^1+\beta \delta x^1,$ 
$ x^2+\gamma \delta x^2,$ 
$\alpha, \beta, \gamma\in [1,0],$
and  is a part of some
$3$-dimensional surface in the spacetime.
Let the orientation of that surface is specified 
by the basic vector $e_{(3)}.$ 
As configuration space is associated with a reference frame,
$x^0$ is temporal coordinate.
For definiteness, let $x^0=t.$
Denote this  cell by $C_{012}.$
Its smallness means that the dependence of the phase density
and of the current density on 
$x^0,$ $x^1$ $x^2$ can be neglected within the cell
with  an accuracy up to terms  of higher order of smallness.

According to definition of the current density,
the quantity of charge passing through the cell $C_{012}$
in the direction $\delta x_{(3)}$ is equal to  
$$
\int\limits_{C_{012}}J=
-J(\delta x^0e_{(0)},\delta x^1e_{(1)},\delta x^2e_{(2)})=
-J_{012}\delta x^0\delta x^1\delta x^2.
$$
The minus sign appears here
in accordance with the integration rule \cite{drivmono}
because orientation of the cell
is specified by the vector $e_{(3)},$
and the set of the vectors  $\{e_{(3)},$  $e_{(0)},$ $e_{(1)},$ $e_{(2)}\}$
is negatively oriented.
 
Let us express this quantity through the phase density.
For given $\delta t,$
particles passing through the cell $C_{012}$ 
have the coordinate $x^3$ differing 
from the coordinate $x^3$ of the cell
less then $(dx^3/dt) \delta t.$
Consider the four-dimensional cell spanned by the edges vectors
$\delta x^0e_{(0)},$ $\delta x^1e_{(1)},$ $\delta x^2e_{(2)},(dx^3/dt)\delta t\, e_{(3)}.$
Denote it by $C_{0123}.$
The cell $C_{012}$ is one of its  faces.
Not all of particles passing through $C_{0123}$
have world lines that cross
the $3-$ dimensional cell $C_{012},$
because their velocity have nonvanishing   components
$dx^0/dt,$ $dx^1/dt,$ and $dx^2/dt,$
which can be regarded as longitudinal
relative to the cell $C_{012}.$  
Therefore, their world lines
may cross  other faces of the cell  $C_{0123},$
which can be considered as flank faces.  
But under assumption that the phase density 
varies slowly, 
for every particle leaving the $4-$dimensional  cell
through flank boundary,
there exists a particle entering the $4-$dimensional  cell
through opposite flank boundary with the same velocity.
Therefore number of particles crossing the face $C_{012}$ 
is equal to number of particles in the cell $C_{012}.$

In order to calculate their number,
consider number of particle in the $6-$dimensional  cell in the phase space
with edge vectors $\delta x^1e_{(1)},$ $\delta x^2e_{(2)},$
$(dx^3/dt)\delta t\, e_{(3)},$
$\delta q_{(4)},$ $\delta q_{(5)},$ $\delta q_{(6)}:$
$$
n_{123456}(t,q)
dx^1\wedge dx^2\wedge dx^3\wedge dq^4 \wedge dq^5\wedge dq^6
(\delta x^1e_{(1)},\delta x^2e_{(2)},(dx^3/dt)\,\delta t e_{(3)},
$$
$$\delta q_{(4)},\delta q_{(5)},\delta q_{(6)})=
n_{123456}(t,q)\delta x^0\delta x^1\delta x^2 v^3 dq^4 \wedge dq^5\wedge dq^6
(\delta q_{(4)},\delta q_{(5)},\delta q_{(6)})
$$
Here $\delta q_{(4)},$ $\delta q_{(5)},$ $\delta q_{(6)}$
are some linearly independent vectors
with vanishing spatial components.
Integrating over all admissible values of the phase coordinates $q^4, q^5, q^6,$
we get the equality
\equation
J_{012}=-\int\limits_{\Omega(x^1,x^2,x^3)}n_{123456}(t,q)v^3(t,q)
\,dq^4 \wedge dq^5\wedge dq^6.
\label{eq:J012}
\endequation
Analogously, we have
$$
J_{013}=\int\limits_{\Omega(x^1,x^2,x^3)}n(t,q)_{123456}v^2(t,q)
\,dq^4 \wedge dq^5\wedge dq^6,
$$
$$
J_{023}=-\int\limits_{\Omega(x^1,x^2,x^3)}n(t,q)_{123456}v^1(t,q)
\,dq^4 \wedge dq^5\wedge dq^6.
$$

\section{Particle distributions for charged particle beam}

To find analytical solutions
of the Vlasov equation 
is very complicated problem.
It arises in charged particle beam physics
for high current beam.
The solutions of the Vlasov equation
for high current beam 
are often called 
self-consistent distributions,
because particles move in the field that is produced by them.
A lot of papers are devoted to 
self-consistent distributions.

We give here two   examples
of stationary particle distributions in the phase space
for charged particle beam,
the Brillouin flow \cite{brill},
and the Kapchinsky-Vladimirsky distribution\cite{kapmono}.
They are degenerate distributions,
because dimension of support of them
is less then dimension of the phase space.
The stationarity means
that the phase density does not depend of the time.

In both cases,
consider nonrelativistic uniformly charged cylindrical beam
in uniform longitudinal magnetic field $H=(0,0,H_z).$
Self field can be found from the Poisson equation
$$
\Delta \varphi=-{e\varrho_0}/{\varepsilon_0}.
$$
Here $\varrho_0$ is spatial density of the particles
inside the beam cross-section. 

Assume that all particles have the same longitudinal velocity.
Transverse particle motion is described by equations
(\ref{eq:partdynamics2}) and (\ref{eq:emforce})
where $E_r=e\varrho_0r/(2\varepsilon_0),$ $E_{\varphi}=E_z=0,$
$B_{r\varphi}=\mu_0rH_z,$ $B_{rz}=B_{\varphi z}=0.$
Integrating these equations, 
we get following integrals of 
transverse motion 
[8-13].
\equation 
M=r^2(\dot\varphi+\omega_0),
\label{eq:M}
\endequation 
\equation
H=\dot r^2+\omega^2r^2+M^2/r^2.
\label{eq:H}
\endequation
Here
$r,\varphi,z$ are cylindrical coordinates,
overdot denotes differentiation with respect to $t,$
$\omega_0=eB_z/(2m),$ $\omega^2=\omega_0^2-e\varrho_0/(m\varepsilon_0)={\rm const},$
$e, m$ are charge and mass of the particle,
$B_z=B_{r\varphi}/r={\rm const}.$

The Brillouin flow is trivial example,
and can be described without concerning of the phase density.
But we apply the approach developed here 
to show how it works.
In the Brillouin flow
all particles rotate around the beam axis
with the same angular velocity.
It is easy to find this velocity and the particle spatial density:
$\dot\varphi=-\omega_0,$
$\varrho=m\omega_0^2\varepsilon_0/e.$
That means that for all particles $M=0,$ and $H=0.$

In order to introduce the phase space,
consider a transverse slice of the beam
moving along the axis:
$z\in (z_0+v^zt,z_0+\delta z+v^zt),$
$\delta z$ being assumed small enough
to consider that all particles of the slice
have the same value of the coordinate $z.$
As assumed previously,
all particles of the slice
have also the same longitudinal component of the velocity.
Therefore, the phase space for the particles of this slice
is four-dimensional in the general case.
But in the case of the Brillouin flow,
all the particles of the slice
lie on the surface $M=0,$ and $H=0.$
Let us take this surface as the phase space.
This phase space is two-dimensional.
Spatial coordinates $r$  and $\varphi$ can be taken as coordinates
also in the phase space,
$r<R,$ where $R$ is the beam radius.
Then the Vlasov equation (\ref{eq:master6}) takes the form
$$
\frac{\partial n_{r\varphi}}{\partial t}+
\dot r\frac{\partial n_{r\varphi}}{\partial r}+
\dot \varphi\frac{\partial n_{r\varphi}}{\partial \varphi}=0.
$$ 
This equation is satisfied,
because all terms 
are equal to $0.$
 
Consider the Kapchinsky-Vladimirsky distribution.
It is also called the microcanonical distribution,
because all the particles   have the same value of the energy 
of the transverse motion $H=H_0.$
For every admissible value of $M$  
there exists a set 
of particle trajectories,
for which $r$ and $\varphi$
change in accordance with equations (\ref{eq:M}), (\ref{eq:H}).
Rotation of some trajectory
by an arbitrary angle around the beam axis gives
another trajectory with the same $M.$
Assume that the particles are evenly distributed
on all the trajectories
with the same $M$ differing by the rotation angle. 

Denote the phase of the particle on the trajectory
by $\theta,$ 
and take $M,$ $\theta$ and $\varphi$
as  coordinates in the phase space,
which is the surface $H=H_0.$
Then the Vlasov equation
takes the form
\equation
\frac{\partial n_{\varphi\theta M}}{\partial t}+
\dot \varphi\frac{\partial n_{\varphi\theta M}}{\partial \varphi}+
\dot \theta\frac{\partial n_{\varphi\theta M}}{\partial \theta}+
\dot M\frac{\partial n_{\varphi\theta M}}{\partial M}=0.
\label{eq:v3}
\endequation
The first term is equal to $0$
according to the stationarity of the distribution.
The second term is  equal to $0$
according to the above assumption
about uniformity of azimuthal distribution.
The forth term is equal to $0$ as $M$ conserves.
Therefore, we obtain 
from the Vlasov equation 
that the  particles should be evenly distributed
on phases $\theta$ of the trajectories, as $\dot \theta\neq 0.$.
 
The Vlasov equation is satisfied,
and the problem seems to be solved,
but it is more complicated 
because the solution is obtained 
under assumption that beam is uniformly charged.
Let us show that the spatial density of the beam is uniform
if the  particles are uniformly distributed on $M$
on a segment
 $(-M_0,M_0).$

First of all, note that beam boundary $r=R$
can be reached only by particles with $M=0.$ 
As $\dot r=0$ at $r=R,$
we have that $H_0=\omega^2R^2.$

It can be seen from (\ref{eq:H}) that 
radial velocity changes as 
\equation
\dot r=\pm\sqrt{H-\omega^2r^2-\frac{M^2}{r^2}} 
\label{eq:vr}
\endequation
along a particle trajectory.
It is easy to get that maximal value of $|M|$ 
is reached when the particle moves along circular trajectory: $\dot r=0.$
For circular trajectory $r=\sqrt{|M|/\omega},$
and therefore $H=2\omega|M|.$
Taking into account that $H=\omega^2R^2,$
we get that maximal value of $|M|$ is $M_0=\omega R^2/2.$

Let us introduce other coordinates in the phase space:
$x,$ $y,$ $M$
where $x$ and $y$ are Cartesian coordinates in the configuration space.
Components $n_{xyM}$ and $n_{\phi\theta M}$
are related by the tensor component transformation rule,
which in this case can be written as
$$
n_{xyM}=2n_{\varphi\theta M}\cdot\det
\biggl\vert
\begin{pmatrix}
\frac{\partial\varphi}{\partial x}&\frac{\partial\varphi}{\partial y}\cr
\frac{\partial\theta}{\partial x}&\frac{\partial\theta}{\partial y}
\end{pmatrix}
\biggr\vert
=\frac{2n_{\varphi\theta M}}{r|\dot r|}.
$$
Here we take into account that
each trajectory passes twice through
every point in the phase space that lies on it,
when the particle approach to the axis,
and when the particle move away from the axis.
These cases differ only by the sign of the radial component of velocity.
The multiplier $2$ is written in compliance with this reason.

Then 
$$
\varrho=
\int\limits_{-M_0}^{M^0}n_{xyM}\,dM=
\frac{2n_{\varphi\theta M}}r
\int\limits_{-M_0}^{M_0}
\frac{dM}{(H_0-M^2/r^2-\omega^2r^2)^{1/2}}=2\pi n_{\varphi\theta M}={\rm const}.
$$
Thus, we have proved that  spatial density is uniform
inside the beam cross-section.
That means that uniform phase density
defined on the surface $H=H_0,$ $M\in(-M_0,M_0)$
corresponds to the Kapchinsky-Vladimirsky distribution.

\section{Conclusion}

The covariant theory
of the phase space distributions
for particles moving in the spacetime is presented.
If there are introduced coordinates in the spacetime,
it is possible to consider a phase space.
It turns out that a phase space is not
necessarily associated with a reference frame.
But from physical point of view, 
it is preferable to use phase space
associated with a reference frame .
As it is shown in the work \cite{refframes,refframes2}
the reference frame can be defined by the same manner
both in relativistic and nonrelativistic cases.
Therefore, the concept  of  phase space
associated with a reference frame
is the same for both cases.
The differences between both cases are related only 
to the way of specifying of the metric tensor
and to the form of motion equations.

The approach presented here
does not use the concept of phase volume,
as compared to the common approach.
If one introduces the phase volume,
then particle distribution function
$\overline n$ can be defined as multiplier
in the equality $n={\overline n}\Omega_P$ \cite{groot,plasmas2013}
where $n$ is the phase  density form of top degree and $\Omega_P$
is the phase volume form 
(the Liouville 6-form in Ref.\cite{plasmas2013}).

Firstly, that approach requires rigorous definition
of the phase volume form.
In Ref. \cite{plasmas2013}
the Liouville 6-form
is defined in the 7-dimensional space,
and reduction to the 6-dimensional
space is carried out only
for the case of the flat Minkowski spacetime.
In the present approach
the phase space is 6-dimensional,
and the 7-dimensional space is required
only for computation of self force 
using the Maxwell equations that contain 
4-dimensional electric current density.

Secondly, the definition of the phase density
as the distribution function can be used
only in the case of nondegenerate distribution,
when particles are distributed 
in some open subdomain of the phase space.
By this reason, such approach  faces difficulties
for degenerate distributions, when particles are distributed on
some surface in the phase space. 

As it is shown by the examples,
the   approach presented here is simple in use,
and plainer, because it does not contain
unnecessary conception of the phase volume.

The  theory presented here is rigorous basis
for consideration of particle distributions
for a charged particle beam,
particularly,  of degenerate distributions.
Degenerate particle distributions
are often used as model distributions,
for example, for numerical solutions 
of the optimization problems
for charged particle accelerators
\cite{drstar}.

The theory can be also applied in other problems
concerned with self-consistent field 
of moving particles,
for example, in the problem
of self-gravitating star matter.

\end{document}